\documentclass[aps,prb,twocolumn,showpacs,floatfix,superscriptaddress]{revtex4-1}
\usepackage[dvips]{graphics}
\usepackage{graphicx}
\usepackage[utf8]{inputenc}
\usepackage{amssymb}
\usepackage{epstopdf}
\usepackage{bm}

\begin{document}
\title{Signature of the long range triplet proximity effect in the density of states}
\author{Milo\v s Kne\v zevi\'c}
\affiliation{Department of Physics, University of Belgrade, P.O. Box 368, 11001
Belgrade, Serbia}
\affiliation{Cavendish Laboratory, University of Cambridge, JJ Thomson Av., Cambridge CB3 0HE, UK}
\author{Luka Trifunovic}
\affiliation{Department of Physics, University of Belgrade, P.O. Box 368, 11001
Belgrade, Serbia}
\affiliation{Department of Physics, University of Basel, Klingelbergstrasse 82, CH-4056 Basel, Switzerland}
\author{Zoran Radovi\'c}
\affiliation{Department of Physics, University of Belgrade, P.O. Box 368, 11001
Belgrade, Serbia}

\begin{abstract}
  We study the impact of the long-range spin-triplet proximity effect on the density of
  states (DOS) in planar SF$_1$F$_2$S Josephson junctions that consist of conventional superconductors (S) connected
  by two metallic monodomain ferromagnets (F$_1$ and F$_2$) with transparent
  interfaces. We determine the electronic DOS in F layers and the Josephson current
  for arbitrary orientation of the magnetizations using the solutions of Eilenberger equations in
  the clean limit and for a moderate disorder in ferromagnets.
  We find that fully developed long-range proximity effect can occur in highly asymmetric ferromagnetic bilayer Josephson junctions with orthogonal magnetizations. The effect manifests itself as an enhancement in DOS, and as a dominant second harmonic in the Josephson current-phase relation.
  Distinctive variation of DOS in ferromagnets with the angle between magnetizations
  is experimentally observable by tunneling spectroscopy. 
  This can provide an unambiguous signature of the long-range spin-triplet proximity effect.
\end{abstract}


\maketitle

\section{Introduction}
Long-range spin-triplet superconducting correlations induced in heterostructures comprised
of superconductors with the usual singlet pairing and inhomogeneous
ferromagnets have been predicted recently.~\cite{bergeret_long-range_2001,volkov_odd_2003,bergeret_odd_2005}
It has been experimentally verified that superconducting correlations can
propagate from SF interfaces with a penetration length up to 1$\mu$m,
and provide a nonvanishing Josephson supercurrent
through very strong ferromagnets.~\cite{sosnin_superconducting_2006,keizer_spin_2006,anwar_supercurrents_2010,wang_interplay_2010,
sprungmann_evidence_2010,khaire_observation_2010,robinson_controlled_2010}

The simplest Josephson junction with inhomogeneous
magnetization is SF$_1$F$_2$S structure with monodomain
ferromagnetic layers having noncollinear in-plane magnetizations.
However, the long-range Josephson effect is not feasible in the
junctions with only two F layers,\cite{braude_fully_2007,houzet_long_2007,volkov_odd_2010,pajovi_josephson_2006,crouzy_josephson_2007}
except in highly asymmetric
SF$_1$F$_2$S junctions at low temperatures, as it was
shown in Refs.~\onlinecite{trifunovic_josephson_2011} and ~\onlinecite{TrifunovicPRL107-11}. In that case, the long-range spin-triplet
effect manifests itself as a large second harmonic ($I_2\gg I_1$)
in the expansion of the Josephson current-phase
relation, $I(\phi)=I_1 \sin(\phi)+I_2 \sin(2\phi)+\cdots$.
The ground state in Josephson junctions with ferromagnetic barriers can be either $0$ or $\pi$
state.
The energy of the junction is proportional to
$\int_0^{\phi}{I(\phi')d\phi'}$, hence a second harmonic leads to
degenerate ground states at $\phi=0$ and $\phi=\pi$.
Small contribution of the first harmonic lifts the degeneracy
which results in coexistence of stable and metastable $0$ and
$\pi$ states.\cite{radovi_coexistence_2001}

In SFS Josephson junctions with
homogeneous magnetization, the projection
of the total spin of a pair to the direction of magnetization is
conserved and only spin-singlet and triplet correlations with
zero spin projection occur.~\cite{buzdin_proximity_2005} These
correlations penetrate into the ferromagnet over a short distance
(determined by the exchange energy $h$)
$\xi_F=\hbar v_F/h$ in the clean limit
and $\xi_F=\sqrt{\hbar D/h}$ in the dirty limit,
where diffusion coefficient $D=v_F l/3$.
For inhomogeneous magnetization,
odd-frequency triplet correlations with nonzero ($\pm 1$) total spin
projection are present as well. These correlations, not suppressed by the
exchange interaction, are long-ranged, $\xi_F=\hbar v_F/k_BT$ ($\sqrt{\hbar D/k_BT}$) in the clean
(dirty) limit, and have a dramatic impact
on the Josephson effect.~\cite{bergeret_odd_2005}

Dominant influence of long-range triplet correlations on the Josephson current can be
realized in SFS junctions with magnetically active
interfaces,~\cite{eschrig_theory_2003,eschrig_triplet_2008}
narrow domain walls between S and thick F interlayers with misaligned
magnetizations,~\cite{braude_fully_2007,houzet_long_2007,volkov_odd_2008,
volkov_odd_2010,trifunovic_long-range_2010,alidoust_spin-triplet_2010-1} or
superconductors with spin orbit interaction.~\cite{PhysRevB.83.054513}
In addition to the impact on the Josephson current, the odd-frequency triplet
pair correlations can be seen through enhanced electronic density of states by tunneling spectroscopy.\cite{Cottet}
It has been found that proximity effect in diffusive FNS or NFS structures (N is a normal nonmagnetic metal), with magnetically active interface or precessing magnetization, gives a clearcut signature of the odd-frequency superconducting correlations.~\cite{yokoyama_manifestation_2007,yokoyama_tuning_2009}
In general, measuring DOS is a very powerful tool to characterize the nature of superconducting
correlations in NS and FS structures.\cite{BruderBelzigPilgram,EschrigLohneysen}

In this article, we study the proximity effect and influence of
spin-triplet superconducting correlations in clean and moderately diffusive
SF$_1$F$_2$S junctions with transparent interfaces. Magnetic interlayer is composed of two
monodomain ferromagnets with arbitrary orientation of in-plane magnetizations. We
calculate density of states (DOS) in ferromagnetic layers and the Josephson current from the
solutions of the Eilenberger equations.
Previously, F$_1$SF$_2$ and SF$_1$F$_2$S junctions with monodomain
ferromagnetic layers having noncollinear in-plane magnetizations have been studied using Bogoliubov-de Gennes
equation~\cite{pajovi_josephson_2006,asano_josephson_2007,halterman_odd_2007,halterman_induced_2008}
and within the quasiclassical approximation in diffusive~\cite{fominov_triplet_2003,
you_magnetization-orientation_2004,loefwander_interplay_2005,barash_josephson_2002,fominov_josephson_2007,
crouzy_josephson_2007, sperstad_josephson_2008,karminskaya_josephson_2009} and
clean~\cite{blanter_supercurrent_2004,linder_proximity_2009} limits using Usadel
and Elienberger equations, respectively.

We present analytical solutions in stepwise approximation in the
clean limit, and numerical self-consistently obtained results both in the clean limit and for a
moderate disorder in ferromagnets.
The second harmonic in the Josephson current-phase relation is dominant for highly asymmetric SF$_1$F$_2$S junctions composed of particularly thin (weak) and thick (strong) ferromagnetic layers with noncollinear magnetizations.
This is a manifestation
of the long-range spin-triplet proximity effect at low temperatures,
related to the phase coherent transport of two Cooper pairs.~\cite{trifunovic_josephson_2011, TrifunovicPRL107-11}
The proximity effect is also accompanied by pronounced change of both shape and magnitude of electronic DOS in ferromagnetic layers near the Fermi surface. In this case, the self-consistency and finite (but large)
electronic mean free path do not alter the analytical results qualitatively.
On the other hand, in symmetric SFFS junctions,
with equal and strong magnetic influence of F layers,
the long-range proximity effect is weak (cannot be dominant\cite{trifunovic_josephson_2011});
we show that DOS is therefore equal to its normal metal value even for moderately diffusive ferromagnets.

The article is organized as follows. In Sec. II, we present the
model and the equations that we use to calculate the electronic DOS and the Josephson current.
In Sec. III, we provide results for different ferromagnetic layers
thickness and orientation of magnetizations.
The conclusion is given in Sec. IV.

\section{Model and solutions}

\begin{figure}
  \includegraphics[width=7cm]{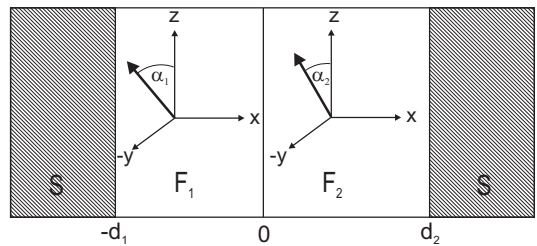}
  \caption{Schematics of an SF$_1$F$_2$S heterojunction. The magnetization
  vectors lie in the y-z plane at angles $\alpha_1$ and $\alpha_2$ with respect
  to the $z$-axis.}
  \label{geometrija}
\end{figure}

We consider a simple model of an SF$_1$F$_2$S  heterojunction consisting of two
conventional ($s$ wave, spin-singlet pairing) superconductors (S) and two uniform monodomain
ferromagnetic layers (F$_1$ and F$_2$) of thickness $d_1$ and $d_2$, with angle
$\alpha=\alpha_2-\alpha_1$ between their in-plane magnetizations (see
Fig.~\ref{geometrija}). Interfaces between layers are fully transparent and
magnetically inactive.

We describe superconductivity in the framework of the Eilenberger
quasiclassical theory.~\cite{bergeret_odd_2005,eilenberger_general_1966} Ferromagnetism
is modeled by the Stoner model, using an exchange energy shift $2h$ between the
spin subbands. Disorder is characterized by the electron mean free path
$l=v_F\tau$, where $\tau$ is the average time between  scattering on impurities,
and $v_F$ is the Fermi velocity assumed to be the same in S and F metals.

Both the clean and moderately diffusive ferromagnetic layers are considered.
In the clean limit, the mean free path $l$ is larger than the two characteristic
lengths: the ferromagnetic exchange length $\xi_F=\hbar v_F/h$, and the
superconducting coherence length $\xi_S=\hbar v_F/\pi\Delta_0$, where $\Delta_0$
is the bulk superconducting pair potential. For moderate disorder,
$\xi_F<l<\xi_S$.

In this model, the Eilenberger Green functions
$g_{\sigma\sigma'}(x,\theta,\omega_n)$,
$g^{\dag}_{\sigma\sigma'}(x,\theta,\omega_n)$,
$f_{\sigma\sigma'}(x,\theta,\omega_n)$, and
$f^{\dag}_{\sigma\sigma'}(x,\theta,\omega_n)$ depend on the Cooper pair center-of-mass
coordinate $x$ along the junction axis, angle $\theta$ of the quasiclassical
trajectories with respect to the $x$ axis,
and on the Matsubara frequencies $\omega_n=\pi k_BT(2n+1)$,
$n=0,\pm1,\ldots$. Spin indices are $\sigma=\uparrow,\downarrow$.
The Eilenberger equation in particle-hole $\otimes$ spin space can be written in
the compact form
\begin{equation}
  \label{Eil}
  \hbar v_x\partial_x \check{g}+\Big[\omega_n \hat{\tau}_3\otimes \hat{1}-i
  \check{V}-\check{\Delta}+\hbar \langle \check{g}\rangle/2\tau,
  \check{g}\Big]=0,
\end{equation}
with normalization condition $\check{g}^2=\check{1}$. We indicate by
$\hat{\cdots}$ and $\check{\cdots}$ $2\times2$ and $4\times4$ matrices,
respectively. The brackets $\langle ...\rangle=(1/2)\int_0^{\pi}(...)\sin (\theta)d \theta$
denote angular averaging over the Fermi surface, $[\;,\;]$ denotes a
commutator, and $v_x=v_F\cos{\theta}$ is the projection of the Fermi velocity vector
on the $x$ axis.

The matrix of quasiclassical Green functions is\cite{trifunovic_josephson_2011,fUnitary}
%
%\footnote{Our definition of matrix $\check g$ insignificantly differs from the choice in Ref.~\onlinecite{bergeret_odd_2005} by an unitary transformation.}
%
\begin{equation}
  \check{g}=\left[
   \begin{array}{rrrr}
      g_{\uparrow\uparrow} & g_{\uparrow\downarrow} &
      f_{\uparrow\uparrow} & f_{\uparrow\downarrow}\\ g_{\downarrow\uparrow} &
      g_{\downarrow\downarrow} & f_{\downarrow\uparrow} &
      f_{\downarrow\downarrow}\\ -f_{\uparrow\uparrow}^{\dag} &
      -f_{\uparrow\downarrow}^{\dag} & -g_{\uparrow\uparrow}^{\dag} &
      -g_{\uparrow\downarrow}^{\dag}\\ -f_{\downarrow\uparrow}^{\dag} &
      -f_{\downarrow\downarrow}^{\dag} & -g_{\downarrow\uparrow}^{\dag} &
      -g_{\downarrow\downarrow}^{\dag}
    \end{array}  \right],
\end{equation}
and the matrix $\check{V}$ is given by
\begin{eqnarray}
  \check{V}=\hat{1}\otimes \textrm{Re}\Big[
  \textbf{h}(x)\cdot\widehat{\bm{\sigma}}\Big]+i\hat{\tau}_3\otimes
  \textrm{Im}\Big[ \textbf{h}(x)\cdot\widehat{\bm{\sigma}}\Big],
\end{eqnarray}
where the components  $\hat{\sigma}_x, \hat{\sigma}_y, \hat{\sigma}_z$ of the
vector $\widehat{\bm{\sigma}}$, and $\hat{\tau}_1, \hat{\tau}_2, \hat{\tau}_3$
are the Pauli matrices in the spin and the particle-hole space, respectively.
The in-plane ($y$-$z$) magnetizations of the neighboring F layers are not
collinear in general, and form angles $\alpha_1$ and $\alpha_2$ with respect to
the $z$-axis in the left (F$_1$) and the right (F$_2$) ferromagnets.
The exchange field in ferromagnetic layers is
$\textbf{h}(x)=h_{1}(0,\sin{\alpha_1},\cos{\alpha_1})$ and
$h_{2}(0,\sin{\alpha_2},\cos{\alpha_2})$.

We assume that the superconductors are identical, with
\begin{equation}
  \check{\Delta}=\left[ \begin{array}{cc} 0 & \hat{\sigma}_2\Delta\\
    \hat{\sigma}_2\Delta^{*} & 0 \end{array}  \right]
\end{equation}
for $x<-d_1$ and $x>d_2$.
The self-consistency condition for the pair potential
$\Delta=\Delta(x)$ is given by
\begin{eqnarray}
  \label{samo}
  \Delta=-i\lambda \pi N_0(0) k_BT\sum_{\omega_n}\langle
  f_{\uparrow\downarrow}\rangle,
\end{eqnarray}
where $\lambda$ is the coupling constant, $N_0(0)=mk_F/\pi^2\hbar^2$ is the
density of states (per unit volume) of the free electron gas at the Fermi level
$E_F=\hbar^2 k_F^2/2m$, and $k_F=mv_F/\hbar$ is
the Fermi wave number. In F layers $\check{\Delta}=0$.

The density of states normalized by its normal-state value is given by analytical continuation
\begin{equation}
  \label{dos}
  N(\varepsilon)/N_0(0) = \frac{1}{2}\; \sum_{\sigma} \langle \textrm{Re}\hspace{1mm} g_{\sigma \sigma}(i \omega_n \to \varepsilon + i \delta) \rangle,
\end{equation}
where a small imaginary part of the energy is introduced to regularize singularities. In numerical calculations we choose
$\delta/\Delta_0(0) = 10^{-2}$.

The supercurrent is obtained from the normal Green function through the following
expression
\begin{equation}
  \label{struja}
  I(\phi)=\frac{1}{2} \pi e N_0(0)Sk_BT\sum_{\omega_n}\sum_{\sigma}\langle v_x
  \textrm{Im}\hspace{1mm} g_{\sigma\sigma}\rangle ,
\end{equation}
where $\phi$ is the macroscopic phase difference across the junction, and $S$ is
the area of the junction. In our examples, the current is normalized to
$\pi \Delta_0/eR_N$ where $R_N=2\pi^2\hbar/Se^2k_F^2$.

We consider only transparent interfaces and use continuity of the Green
functions as the boundary conditions.
Analytical solutions of Eq.~(\ref{Eil}) are derived (see Appendix) in the stepwise approximation for the pair potential
\begin{equation}
  \label{del}
  \Delta=\Delta_0 \left[ e^{-i\phi /2}\Theta
  (-x-d_{1})+e^{i\phi /2}\Theta (x-d_{2})\right],
\end{equation}
where $\Delta_0$ is the bulk pair potential, and $\Theta(x)$ is the Heaviside step function.
The temperature dependence of the bulk pair potential $\Delta_0 $ is given
by $\Delta_0 (T)=\Delta_0 (0)\tanh \left(
1.74\sqrt{T_{c}/T-1}\right)$.~\cite{muehlschlegel_thermodynamischen_1959}

We have chosen $\alpha_2 =0$, that is $\alpha = \alpha_1$. For three characteristic values of the angle between magnetizations $\alpha=0$, $\pi/2$, and $\pi$, the normal Green functions in F$_2$ layer are $x$-independent and can be written in a compact form
\begin{widetext}
\begin{equation}
  \label{eq9}
  g_{\sigma \sigma}(\theta, \omega_n) = \frac{\omega_n}{\Omega_n} + \frac{\Delta_0 ^2}{\Omega_n} \;
  \frac{1}{\omega_n + i \Omega_n \hspace{1mm}\hbox{sign}(\cos{\theta})\cot{\left (\pm \frac{\vartheta_1 + \vartheta_2}{2} + i \frac{\Xi_n}{2} - \frac{\phi}{2} \right )}},
\end{equation}
\begin{equation}
  \label{eq10}
  g_{\sigma \sigma}(\theta, \omega_n) = \frac{\omega_n}{\Omega_n} + \frac{\Delta_0^2}{\Omega_n} \;
  \frac{\omega_n [\cos{(\phi - i\Xi_n)} - \cos{(\vartheta_1)} \cos{(\vartheta_2)}] - i \Omega_n \hspace{1mm}\hbox{sign}(\cos{\theta}) [\pm \cos{(\vartheta_1)} \sin{(\vartheta_2)} - \sin{(\phi - i\Xi_n)}]}{(\Omega_n ^2 + \omega_n ^2)\cos{(\phi - i\Xi_n)} + (\Omega_n ^2 - \omega_n ^2) \cos{(\vartheta_1)} \cos{(\vartheta_2)} + 2i\omega_n \Omega_n \hspace{1mm}\hbox{sign}(\cos{\theta}) \sin{(\phi - i\Xi_n)}},
\end{equation}
\begin{equation}
  \label{eq11}
  g_{\sigma \sigma}(\theta, \omega_n) = \frac{\omega_n}{\Omega_n} + \frac{\Delta_0^2}{\Omega_n} \;
  \frac{1}{\omega_n + i\Omega_n \hspace{1mm}\hbox{sign}(\cos{\theta})\cot{\left ( \pm \frac{\vartheta_2 - \vartheta_1}{2} + i \frac{\Xi_n}{2} - \frac{\phi}{2} \right )}},
\end{equation}
\end{widetext}
respectively. Here, $\pm$ corresponds to $\sigma = \uparrow, \downarrow$, $\Omega_n^2 = \omega_n^2+\Delta_0^2$, $\Xi_n = 2(d_1+d_2)\omega_n/\hbar v_F \cos{\theta}$ and $\vartheta_j = 2 d_j h_j/\hbar v_F \cos{\theta}$, $j=1$, $2$.

The well-known normal Green functions for $\alpha = 0$ and $\pi$ are the same in F${}_1$ and F$_2$ layers, Eqs.~(\ref{eq9}) and~(\ref{eq11}).\cite{konschelle_nonsinusoidal_2008, blanter_supercurrent_2004}
However, for $\alpha = \pi/2$ only $\sum_{\sigma} g_{\sigma \sigma}$ is the same in F$_1$ and F$_2$.
In F$_1$ layer, $g_{\sigma \sigma}$ is $x$- and $\sigma$-dependent and cannot be expressed in a compact form.

The energy of Andreev states is determined by the poles of the Green functions. In contrast to the case of collinear magnetizations,
the spin splitting of Andreev states is absent for orthogonal magnetizations, Eq.~(\ref{eq10}), where the long-range spin-triplet proximity effect is maximally pronounced.

Real part of the Green functions as a function of energy $\varepsilon$ inside the superconducting gap, $-\Delta_0 < \varepsilon < \Delta_0$, reduces to four delta functions with $\theta$-dependent position for collinear magnetizations, and two delta functions
for orthogonal magentizations. For planar SF$_1$F$_2$S junctions these contributions should be numerically summed over $\theta$. Note that the electronic DOS in ferromagnets is independent of $x$ only in the clean limit.
For a finite electronic mean free path, DOS is calculated in the middle of F${}_2$ ferromagnetic layer.

The above results for DOS in the clean limit, obtained in the stepwise approximation for the pair potential,
are not altered qualitatively in numerical self-consistent calculations.
Numerical computation for both the clean limit and for moderately diffusive ferromagnets is carried out using the collocation method: Eq.~(\ref{Eil}) is solved  iteratively together with the self-consistency condition, Eq.~(\ref{samo}).
Iterations are performed until self-consistency is reached, starting from the
stepwise approximation for the pair potential.
For a finite electron mean free
path in ferromagnets, the iterative procedure starts from the clean limit. We
choose the appropriate boundary conditions in
superconductors at the distance exceeding $2\xi_S$ from the SF interfaces.
These boundary conditions are determined by eliminating the unknown constants from the
analytical solutions in stepwise approximation. To reach self-consistency with good accuracy,
starting from the stepwise $\Delta$, five to ten iterative steps were sufficient.

\section{Results}

\begin{figure}
  \includegraphics[width=9cm]{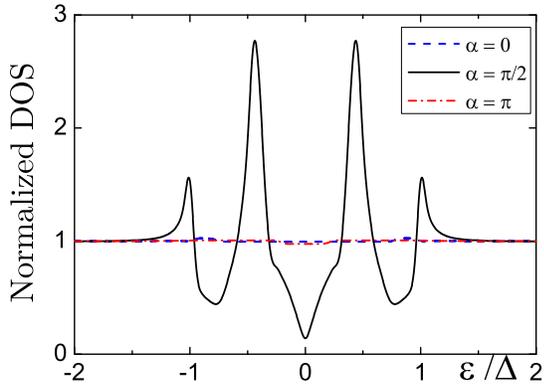}
  \caption{(Color online) Normalized DOS, $N(\varepsilon)/N_0(0)$, in ferromagnets for an asymmetric SF$_1$F$_2$S junction with $d_1=10k_F^{-1}$ and $d_2=990k_F^{-1}$, $h_{1}=h_{2}=0.1E_F$.
  Superconductor is characterized by $\Delta_0(0) /E_{\rm F}=10^{-3}$; the phase difference $\phi=0$.
  The results of non-self-consistent calculations in the clean limit, $l\to \infty$, are shown for three values of the relative angle between magnetizations: $\alpha=0$ (dashed curve), $\pi/2$ (solid curve), and
  $\pi$ (dash-dotted curve).}
  \label{fig2}
\end{figure}

We illustrate our results on SF$_1$F$_2$S planar junctions with relatively weak
ferromagnets, $h/E_F = 0.1$, and the ferromagnetic exchange length $\xi_F =
20k_F^{-1}$. Superconductors are characterized by the bulk pair potential at
zero temperature $\Delta_0(0) /E_{\rm F}=10^{-3}$, which corresponds to the
superconducting coherence length $\xi_S(0)=636 k_F^{-1}$.
We assume that all interfaces are fully
transparent and the Fermi wave numbers in all metals are equal ($k_F^{-1}\sim$ 1{\AA}).

\begin{figure}
  \includegraphics[width=8cm]{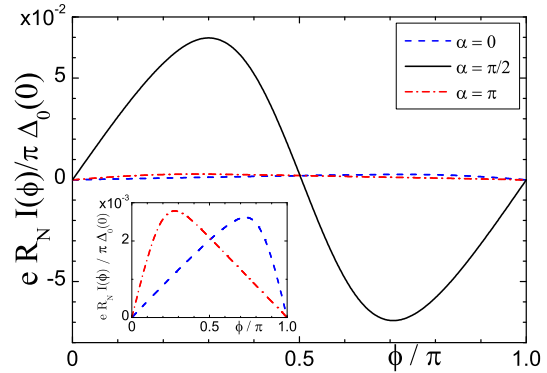}
  \caption{(Color online) The Josephson current-phase relation $I(\phi)$ for an asymmetric SF$_1$F$_2$S
  junction with $d_1=10k_F^{-1}$ and $d_2=990k_F^{-1}$, $h_1=h_2=0.1E_F$,
  $l\to \infty$, $T/T_c=0.1$, and for three values of the relative angle
  between magnetizations: $\alpha=0$ (dashed curve), $\pi/2$ (solid curve), and
  $\pi$ (dash-dotted curve). Inset: Magnified curves for $\alpha=0$ and $\alpha=\pi$. }
  \label{fig3}
\end{figure}

\begin{figure}
  \includegraphics[width=9cm]{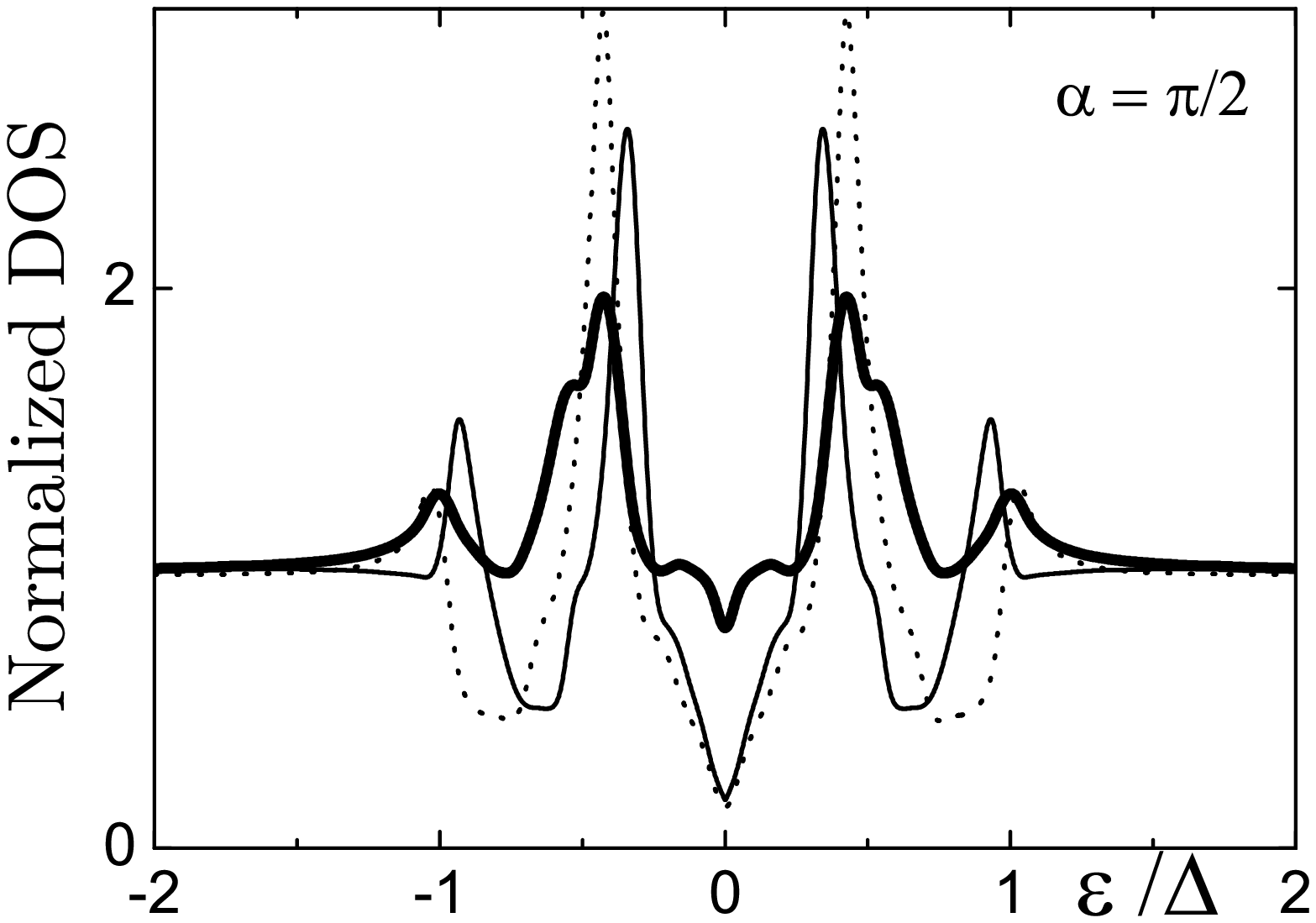}
  \caption{Normalized DOS, $N(\varepsilon)/N_0(0)$, in ferromagnets for an asymmetric SF$_1$F$_2$S junction with $d_1=10k_F^{-1}$ and $d_2=990k_F^{-1}$, $h_{1}=h_{2}=0.1E_F$, for $\alpha=\pi/2$.
  Superconductor is characterized by $\Delta_0(0) /E_{\rm F}=10^{-3}$; the phase difference $\phi=0$.
  Results of self-consistent numerical solutions in the clean limit, $l\to \infty$ (thin solid curve) and for moderately diffusive ferromagnets, $l=200k_F^{-1}$, $x=d_2/2$ (thick solid curve) are shown. For comparison, the non-self-consistent solution in the clean limit (dotted curve) is also shown. }
  \label{fig4}
\end{figure}

\begin{figure}
  \includegraphics[width=9cm]{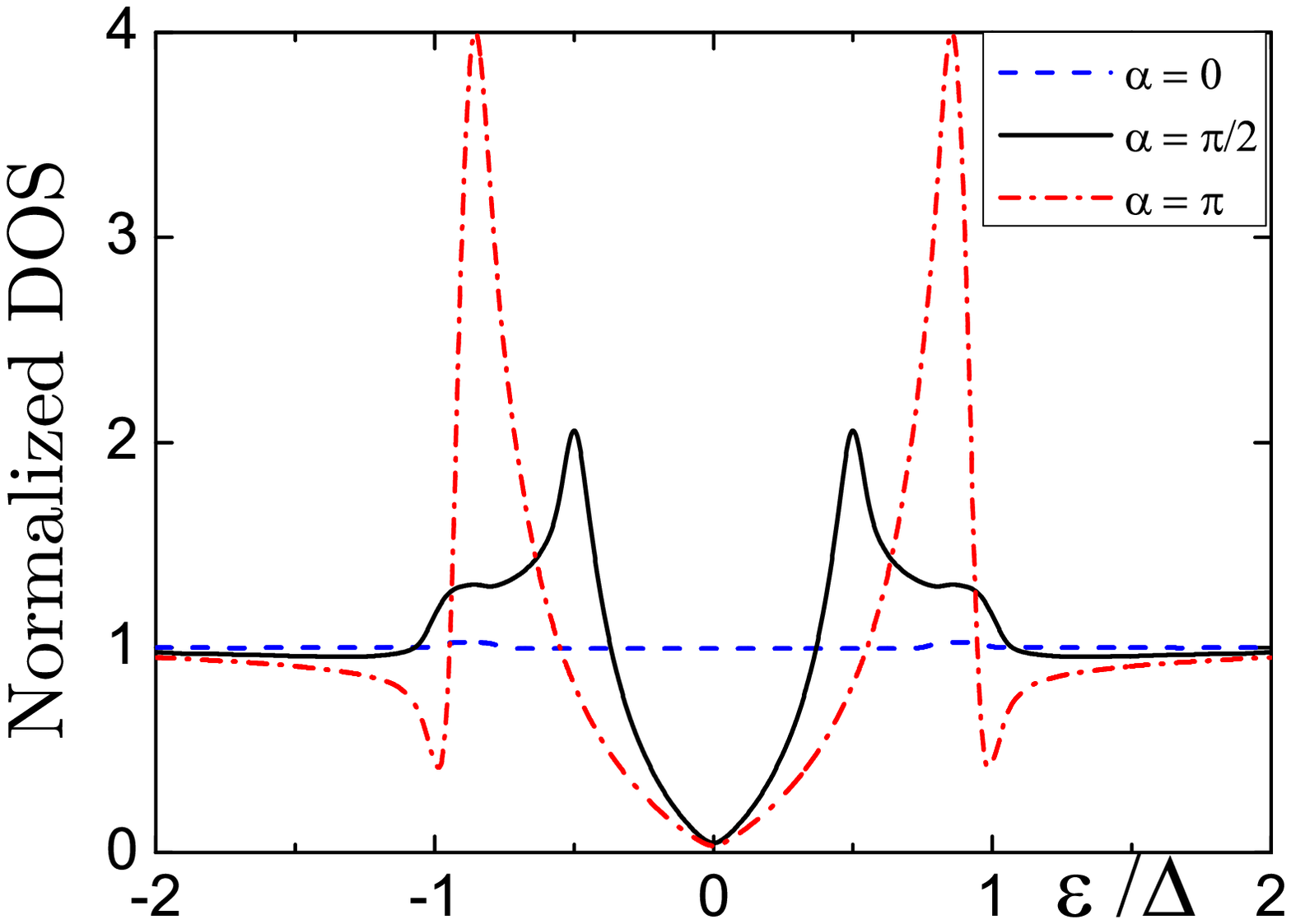} \caption{(Color online) Normalized DOS, $N(\varepsilon)/N_0(0)$, in ferromagnets for a symmetric SF$_1$F$_2$S junction with $d_1= d_2 =500k_F^{-1}$, $h_{1}=h_{2}=0.1E_F$.
  Superconductor is characterized by $\Delta_0(0) /E_{\rm F}=10^{-3}$; the phase difference $\phi=0$.
  Results of non-self-consistent calculations in the clean limit, $l\to \infty$ are shown for three values of the relative angle between magnetizations: $\alpha=0$ (dashed curve), $\pi/2$ (solid curve), and $\pi$ (dash-dotted curve).}
  \label{fig5}
\end{figure}

\begin{figure}
  \includegraphics[width=9cm]{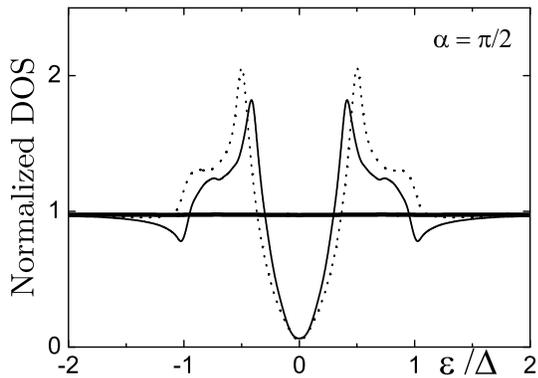}
  \caption{Normalized DOS, $N(\varepsilon)/N_0(0)$, in ferromagnets for a symmetric SF$_1$F$_2$S junction with $d_1= d_2 = 500k_F^{-1}$, $h_{1}=h_{2}=0.1E_F$, for $\alpha=\pi/2$.
  Superconductor is characterized by $\Delta_0(0) /E_{\rm F}=10^{-3}$; the phase difference $\phi=0$.
  Results of self-consistent numerical solutions in the clean limit, $l\to \infty$ (thin solid curve) and for moderately diffusive ferromagnets, $l=200k_F^{-1}$, $x=d_2/2$ (thick solid curve) are shown. For comparison, the non-self-consistent solution in the clean limit (dotted curve) is also shown. }
  \label{fig6}
\end{figure}

Detailed analysis is given for a highly asymmetric junction ($d_1=10k_F^{-1}$ and $d_2=990k_F^{-1}$),
Figs. \ref{fig2}-\ref{fig4}, and for a symmetric junction ($d_1=d_2=500k_F^{-1}$), Figs. \ref{fig5} and \ref{fig6}.
In these examples: $h_1=h_2=0.1E_F$ and $T/T_c=0.1$.
In Figs. \ref{fig2}, \ref{fig3}, and \ref{fig5}, the results of non-self-consistent calculations for DOS and Josephson currents are
shown in the clean limit ($l\rightarrow\infty$). The self-consistent numerical calculations for electronic DOS in the clean limit ($l\rightarrow\infty$) and for moderate disorder ($l = 200 k_F^{-1}$) in ferromagnets are shown in Figs. \ref{fig4} and \ref{fig6}.


In highly asymmetric SF$_1$F$_2$S junctions the variation of DOS
for finite electronic mean free path is less pronounced but qualitatively similar to the clean limit results,
see Fig. \ref{fig4}. This is not the case for symmetric junctions where the long-range
spin-triplet proximity effect is weak, Fig. \ref{fig6}.

Influence of the ferromagnet is predominantly determined by the parameter
$\Theta=(h/E_F)k_F d$, and the results shown here remain the same for a
large range of the junction parameters where $\Theta_1\sim 1$ in F$_1$ and $\Theta_2\sim 100$ in F$_2$ layers.
Note that $\Theta_1 \sim 1$ is the optimal choice for long-range triplet Josephson effect.~\cite{trifunovic_long-range_2010}
For experimental realization of asymmetric junctions it is more suitable to use weak and strong
ferromagnets ($h_1 \ll h_2$) with small and comparable thickness ($d_1 \approx d_2$) to avoid
multidomain magnetic structures and destructive influence of the orbital effect (vortices).~\cite{Pokrovsky}

We have verified by additional calculations that the long-range triplet proximity effect
in highly asymmetric SF$_1$F$_2$S junctions is rather robust.
Small variations in the layers thickness and in the
exchange energy of ferromagnets, as well as moderate
disorder, do not affect distinctive feature of DOS.
Likewise, the long-ranged second
harmonic in $I(\phi)$ at low temperatures remains dominant.\cite{trifunovic_josephson_2011}

In all calculations we have assumed transparent SF interfaces and clean or moderately disordered ferromagnets. Finite transparency, as well as large disorder, strongly suppress the Josephson current and the electronic DOS approaches its normal state value.

In symmetric SFFS junctions with equal ferromagnetic layers no substantial impact of long-range spin-triplet
superconducting correlations on the Josephson current has been found previously both in the ballistic and diffusive
regimes.~\cite{pajovi_josephson_2006,crouzy_josephson_2007,trifunovic_josephson_2011,TrifunovicPRL107-11}
It can be seen from Figs. \ref{fig5} and \ref{fig6} that the electronic DOS for $\alpha = \pi/2$ is much
less pronounced than for the corresponding (equal strength and total thickness of ferromagnets) 
asymmetric junction, Figs. \ref{fig2} and \ref{fig3}. 
Moreover, for moderate disorder in ferromagnts, $l=200 k_F^{-1}$, the proximity effect is practically absent, and DOS equals its normal value (Fig. \ref{fig6}). 
In symmetric SFFS junctions the Josephson critical current monotonically increases with angle between the magnetizations, except in
the vicinity of $0 - \pi$ transitions.~\cite{trifunovic_josephson_2011} In the clean
limit for $\alpha=\pi$, the influence of opposite magnetizations in F layers practically cancels out 
and the proximity effect is much stronger, see Fig. \ref{fig5}. Both DOS
and the Josephson current-phase relation are the same as for the corresponding SNS junction, where N stands for a normal nonmagnetic metal ($h_1=h_2=0$).
Recently, substantially enhanced supercurrent has been observed in Josephson junctions with ferromagnetic
Fe/Cr/Fe trilayer in the antiparallel configuration, without long-range spin-triplet
correlations.~\cite{robinson_enhanced_2010}

\section{Conclusion}

We have studied the spin-triplet proximity effect in SF$_1$F$_2$S planar junctions made of conventional superconductors and two monodomain ferromagnetic layers with arbitrary thickness, strength, and angle between in-plane magnetizations. We have derived analytical expressions for the quasiclassical normal Green functions in the clean limit and the stepwise pair potential approximation (non self-consistent). We have calculated the electronic density of states in ferromagnets, and the Josephson current. In addition, results of the numerical self-consistent calculations for the clean and moderately diffusive ferromagnets are shown for comparison.

The second harmonic in the Josephson current-phase relation is dominant for highly asymmetric SF$_1$F$_2$S junctions composed of particularly thin (weak) and thick (strong) ferromagnetic layers with noncollinear magnetizations.
This is a manifestation of the long-range spin-triplet proximity effect at low temperatures,
related to the phase coherent transport of two Cooper pairs.
We find that the proximity effect is also accompanied by distinctive variation of electronic DOS in ferromagnetic layers as a function of energy near the Fermi surface. This variation is less pronounced in moderately diffusive case, but still detectable by tunneling spectroscopy as a signature of the long-range proximity effect.
The self-consistency and finite (but large) electronic mean free path do not alter the analytical results qualitatively. In contrast, in symmetric junctions with moderately diffusive and thick ferromagnets the long-range proximity effect is absent and DOS is equal to its normal metal value.

In summary, fully developed long-range proximity effect occurs in highly asymmetric SF$_1$F$_2$S junctions with orthogonal magentizations.
Dominant second harmonic in the Josephson current-phase
relation, as well as a distinctive variation of DOS in ferromagnetic layers with the angle between magnetizations, should be experimentally observable
for relatively small interface roughness and relatively clean ferromagnetic layers
at low temperatures.

%\vspace*{-5mm}
\section{Acknowledgment}
%\vspace*{-5mm}

The work was supported by the Serbian Ministry of Science, Project No.~171027.
Z. R. acknowledges Mihajlo Vanevi\'c, Ivan Bo\v zovi\'c, and Marco Aprili for helpful discussions.

\section{Appendix}

In the clean limit, $l \rightarrow \infty$, solutions of Eq.~(\ref
{Eil}) in the step-wise approximation, Eq. (\ref{del}), for the left superconductor ($x<-d_1$) can be written in the
usual form for normal Green functions

\begin{eqnarray}
\label{gS}
g_{\uparrow\uparrow}(x)&=&\frac{\omega_n}{\Omega_n}+D_1e^{\kappa_sx},\\
g_{\uparrow\downarrow}(x)&=&D_2e^{\kappa_sx},\\
g_{\downarrow\uparrow}(x)&=&D_3e^{\kappa_sx},\\
g_{\downarrow\downarrow}(x)&=&\frac{\omega_n}{\Omega_n}+D_4e^{\kappa_sx},
\end{eqnarray}
 and for anomalous Green functions
\begin{eqnarray}
\label{fS}
f_{\uparrow\uparrow}(x)&=&-\frac{2i\Delta_0}{2\omega_n+\hbar v_x\kappa_s}D_2e^{\kappa_sx},\\
f_{\uparrow\downarrow}(x)&=&\frac{i\Delta_0}{\Omega_n}+\frac{2i\Delta_0}{2\omega_n+\hbar v_x\kappa_s}D_1
e^{\kappa_sx},\\
f_{\downarrow\uparrow}(x)&=&-\frac{i\Delta_0}{\Omega_n}
-\frac{2i\Delta_0}{2\omega_n+\hbar v_x\kappa_s}D_4e^{\kappa_sx},\\
f_{\downarrow\downarrow}(x)&=&\frac{2i\Delta_0}{2\omega_n+\hbar v_x\kappa_s}D_3e^{\kappa_sx},
\end{eqnarray}
 with
\begin{equation}
\kappa_s=\frac{2\Omega_n}{\hbar v_F|\cos(\theta)|},
\end{equation}
and $\Omega_n=\sqrt{\omega_n^2+|\Delta|^2}$. For the right
superconductor ($x>d_2$), the solutions retain the same form with
$\kappa_s\longrightarrow -\kappa_s$, $\phi\longrightarrow -\phi$, with a new set of constants
$D'_1,...,D'_4$.

Solutions for the Green functions in  the left ferromagnetic layer F${_1}$,
$-d_1<x<0$, for three values of angle $\alpha_1 = 0$, $\pi/2$, $\pi$
(we have chosen $\alpha_2=0$)
can be written as follows.
For collinear magnetizations $\alpha_1 =0$ and $\pi$,
\begin{eqnarray}
\label{gF0}
g_{\uparrow\uparrow}(x)&=&K_1,\\
g_{\uparrow\downarrow}(x)&=&K_2e^{\pm i\kappa_{0}x},\\
g_{\downarrow\uparrow}(x)&=&K_3e^{\mp i\kappa_{0}x},\\
g_{\downarrow\downarrow}(x)&=&K_4,
\end{eqnarray}
and
\begin{eqnarray}
\label{fF0}
f_{\uparrow\uparrow}(x)&=&C_1e^{i\kappa x},\\
f_{\uparrow\downarrow}(x)&=&C_2e^{i\kappa_{\pm}x},\\
f_{\downarrow\uparrow}(x)&=&C_3e^{i\kappa_{\mp}x},\\
f_{\downarrow\downarrow}(x)&=&C_4e^{i\kappa x},
\end{eqnarray}
where the upper (lower) sign corresponds to $\alpha_1 =0$ ($\alpha_1 =\pi$).
For orthogonal magnetizations, $\alpha_1 = \pi/2$,
\begin{eqnarray}
\label{gFPi2}
g_{\uparrow\uparrow}(x)&=&K_1 + K_2e^{i\kappa_{0}x} + K_3e^{-i\kappa_{0}x},\\
g_{\uparrow\downarrow}(x)&=&K_4 + i K_2e^{i\kappa_{0}x} - i K_3e^{-i\kappa_{0}x},\\
g_{\downarrow\uparrow}(x)&=&-K_4 + i K_2e^{i\kappa_{0}x} - i K_3e^{-i\kappa_{0}x},\\
g_{\downarrow\downarrow}(x)&=&K_1 - K_2e^{i\kappa_{0}x} - K_3e^{-i\kappa_{0}x},
\end{eqnarray}
and
\begin{eqnarray}
\label{fFPi2}
f_{\uparrow\uparrow}(x)&=&C_1e^{i\kappa x} + C_2e^{i\kappa_{+}x} + C_3e^{i\kappa_{-}x},\\
f_{\uparrow\downarrow}(x)&=&C_4e^{i\kappa x} - i C_2e^{i\kappa_{+}x} + i C_3e^{i\kappa_{-}x},\\
f_{\downarrow\uparrow}(x)&=&C_4e^{i\kappa x} + i C_2e^{i\kappa_{+}x} - i C_3e^{i\kappa_{-}x},\\
f_{\downarrow\downarrow}(x)&=&-C_1e^{i\kappa x} + C_2e^{i\kappa_{+}x} + C_3e^{i\kappa_{-}x}.
\end{eqnarray}
%
%for $\alpha_1 = \pi$
%\begin{eqnarray}
%\label{gFPi}
%g_{\uparrow\uparrow}(x)&=&K_1,\\
%g_{\uparrow\downarrow}(x)&=&K_2e^{-i\kappa_{0}x},\\
%g_{\downarrow\uparrow}(x)&=&K_3e^{i\kappa_{0}x},\\
%g_{\downarrow\downarrow}(x)&=&K_4,
%\end{eqnarray}
%and
%\begin{eqnarray}
%\label{fFPi}
%f_{\uparrow\uparrow}(x)&=&C_1e^{i\kappa x},\\
%f_{\uparrow\downarrow}(x)&=&C_2e^{i\kappa_{-}x},\\
%f_{\downarrow\uparrow}(x)&=&C_3e^{i\kappa_{+}x},\\
%f_{\downarrow\downarrow}(x)&=&C_4e^{i\kappa x}.
%\end{eqnarray}
%
\noindent Here,
\begin{eqnarray}
\kappa_0&=&2h_1/\hbar v_x, \\
\kappa&=&2i\omega_n/\hbar v_x,\\
\kappa_{\pm}&=&2(i\omega_n \pm h_1)/\hbar v_x.
\end{eqnarray}
For the right ferromagnetic layer F${_2}$, $0<x<d_2$,
$\alpha_2=0$, the solutions are the same
as for the layer F${}_1$ in the case $\alpha_1=0$,
with $h_1\to h_2$ and a new set of constants
$K'_1,\ldots,K'_4$, $C'_1,\ldots,C'_4$.

The complete solution has $24$ unknown
coefficients: $4+4$ in superconducting electrodes, and
$8+8$ in ferromagnetic layers. Continuity of normal and anomalous Green functions  at the three interfaces provide the necessary set of $24$ equations.

\bibliography{condmat}
\end{document}